\documentclass[a4paper, english]{paper} 
\usepackage{lipsum, babel}
\usepackage{amsmath,amssymb,amsfonts}
\usepackage[utf8x]{inputenc}
\usepackage[margin=2.5cm]{geometry}
\usepackage{graphicx}
\usepackage{todonotes}
\usepackage{xcolor}
\usepackage[colorlinks=true]{hyperref}

\usepackage{tikz}
\usepackage{physics}
\usepackage{usebib}
\usepackage{comment}

\title{Probabilistic versions of Quantum Private Queries}
\author{Silvia Onofri~$^{1}$ and Vittorio Giovannetti~$^{2}$\\
	\small{$^{1}$ Scuola Normale Superiore, Pisa, Italy}, \\ \small{$^{2}$ NEST, Scuola Normale Superiore and Istituto Nanoscienze-CNR, I-56126 Pisa, Italy}}
\begin{document}
	\maketitle
	\section*{Abstract}
	The no-go theorem regarding unconditionally secure Quantum Bit Commitment protocols is a relevant result in quantum cryptography. Such result has been used to prove the impossibility of unconditional security for other protocols, such as Quantum Oblivious Transfer or One-Sided Two Party Computation. In this paper, we formally define two non-deterministic versions of Quantum Private Queries, a protocol addressing the Symmetric-Private Information Retrieval problem. We show that the strongest variant of such scheme is formally equivalent to  Quantum Bit Commitment, Quantum Oblivious Transfer and One-Sided Two Party Computation protocols. This equivalence serves as conclusive evidence of the impracticality of achieving unconditionally secure Strong Probabilistic Quantum Private Queries.
	
	\section{Introduction}
	The idea of quantum cryptography was introduced by Wiesner in the 1970s and later formalized in the first quantum cryptographic protocol published in 1983 \cite{wiesner-1983}. The design of protocols for cryptography using quantum mechanics looked promising at the beginning: the combination of the no-cloning theorem and the effects of measurements in quantum mechanics create a favorable environment. In fact, these principles make it difficult for an attacker to clone or read an encrypted message without being detected. 
	
	The first protocols of quantum cryptography were then studied with the aim of basing their security exclusively on the principles of quantum mechanics. Some quantum versions of cryptographic primitives such as Quantum Oblivious Transfer \cite{Ardehali_1998}\cite{bennett_brassard_crepeau_skubiszewska}, Quantum Coin Tossing \cite{Brassard_Crepeau} or Quantum Bit Commitment \cite{Ardehali_1996}\cite{Brassard_Crepeau}\cite{brassard_crepeau_jozsa_langlois}
	were published in the early 1990s.
	In particular, a protocol for Quantum Bit Commitment (known as the BCJL protocol) was published in 1993 and declared to be perfectly secure \cite{brassard_crepeau_jozsa_langlois}. Soon after, in 1997, Lo and Chau \cite{Lo_Chau_1997} and, separately, Mayers \cite{Mayers_1996} published counterexamples to the security of the BCJL protocol, showing an efficient attack. After these first results, the same kind of attack was extended to any kind of Quantum Bit Commitment scheme \cite{lo_chau_1998}\cite{mayers_1997}, proving the impossibility of unconditionally secure Quantum Bit Commitment protocols. 
	
	Afterwards, the same proof was extended to other cryptographic primitives, proving the impossibility of protocols such as Quantum Oblivious Transfer, Quantum Coin Tossing, One-Sided Two-Party Computation schemes \cite{lo_1997}. This was an important but negative result: in fact, it was the proof that it is not possible to build such cryptographic primitives by letting their security rely only on the principles of quantum mechanics.
	This paper aims to explore the connections between cryptographic primitives such as Quantum Bit Commitment or One-Sided Two Party Computation with the  Quantum Private Queries (QPQ) protocol published in 2010 by Giovannetti, Lloyd and Maccone \cite{Giovannetti_Lloyd_Maccone_2008, giovannetti_lloyd_maccone_2010}. This protocol addresses the problem of Symmetric-Private Information Retrieval (SPIR), where a user needs to query a database. In its basic form, it provides perfect security on the database side and a good cheating-detection strategy to guarantee user privacy, but in its probabilistic form (i.e., when the database is not deterministic) this detection strategy fails. The connection between this protocol, Quantum Bit Commitment, Quantum Oblivious Transfer and One-Sided Two-Party Computation protocols is shown in the following sections. More precisely, in Sections \ref{subsec:qbc},  \ref{subsec: qot}, \ref{subsec:1s2pc}, and \ref{subsec: qbcand1s2pc} we briefly revise Quantum Bit Commitment, Quantum Oblivious Transfer and One-Sided Two Party Computation protocols and their relations. In Section \ref{sec:QPQ} we introduce the SPIR problem and the Quantum Private Queries protocol, and define two probabilistic versions of it - namely, probabilistic Quantum Private Queries (pQPQ) and Strong probabilistic Quantum Private Queries (SpQPQ). Then, in Section \ref{sec:relations}, we show the connections between the Strong probabilistic version of Quantum Private Queries, Quantum Bit Commitment, Quantum Oblivious Transfer and One-Sided Two Party Computation, and show that the impossibility of the first one follows from the impossibility of the other three protocols.
	
	\subsection{Quantum Bit Commitment (QBC)}
	\label{subsec:qbc}
	Let us briefly introduce what a Quantum Bit Commitment (QBC) scheme is. Bit commitment is a protocol where the sender, Alice, wants to commit to a bit $b=0$ or $b=1$ without immediately revealing its value to the receiver, Bob. Then, the protocol is divided into two main phases: first, in the so-called \textit{commit phase}, Alice chooses the value of $b$, but does not reveal it to Bob; later, in the \textit{opening} phase, she decides to reveal her commitment and Bob discovers the value.
	Then, a secure bit commitment scheme should satisfy two requirements: 
	\begin{itemize}
		\item the protocol should be binding: Alice should be bound to the value of $b$ she chooses at the beginning, i.e., she should not be able to change the value after the commit phase;
		\item the protocol should be concealing: Bob should not be able to determine the value of the bit before the opening phase.
	\end{itemize}
	A general quantum version of this protocol consists in Alice and Bob operating on a Hilbert space $\mathcal{H}=\mathcal A\otimes \mathcal B \otimes\mathcal C$, where $\mathcal A$ and $\mathcal B$ are the quantum private machines of Alice and Bob respectively, while $\mathcal C$ is a public quantum channel.
	In particular, a QBC protocol requires an initial phase (called \textit{preparation-of-states} phase), where Alice should encrypt her choice for $b=0,1$ and prepare her private $\mathcal A$ in an initial state $\ket{0}$ or $\ket{1}$, while $\mathcal B$ and $\mathcal C$ are initialized in a generic state. Then, the commit phase consists of Alice and Bob operating with unitary transformations one on $\mathcal A\otimes \mathcal C$ and one on $\mathcal B\otimes \mathcal C$, and at the end of this phase Bob should have no information about Alice's choice, even if she should already be bound to her commitment. The attack that proved the impossibility of unconditionally secure QBC exploits this point: if Bob has no information about the value of $b$ at the end of the commit phase, Alice can delay her choice at the beginning of the opening phase. More specifically, the fact that Bob has no information about the value of $b$ at the end of the commit phase can be seen as the fact that $\Tr_{\mathcal A}(\ketbra{\psi_0}) =\Tr_{\mathcal A}(\ketbra{\psi_1})$, where $\ket{\psi_0}$ and $\ket{\psi_1}$ are the states of the protocol at the end of the commit phase for $b=0$ and $b=1$, respectively. Because of this equivalence, we can state that $\ket{\psi_0}$ and $\ket{\psi_1}$ have the same Schmidt decomposition and that there exists a unitary transformation, acting only on $\mathcal A$, which takes one to the other. Then, Alice can modify her choice of $b$ until the beginning of the opening phase, so the protocol cannot be both binding and concealing.
	
	\subsection{Quantum Oblivious Transfer (QOT)}
	\label{subsec: qot}
	Oblivious Transfer is another interesting cryptographic primitive, introduced for the first time by Rabin \cite{Rabin_OT}. Since it is more useful for our purposes, we focus on the One-out-of-two Oblivious Transfer (OOT) variant, which is, by the way, fully equivalent to the first one (as proved in \cite{Crepeau_1988}). In OOT, Alice prepares two messages, $m_0$ and $m_1$, and sends them to Bob. Bob can choose only one of them and read it, i.e., he gains full information about the message he chooses, but he learns nothing about the other message. Alice gets no information about which message Bob has chosen. 
	So, the requirements that an OOT protocol should have in order to be secure are:
	\begin{enumerate}
		\item Bob learns the message $m_k$, with $k \in \{0,1\}$;
		\item Alice learns nothing about $k$;
		\item Bob learns nothing about $m_{1-k}$.
	\end{enumerate}
	
	OOT is also equivalent to the so-called One-out-of-$n$ OT (\cite{Brassard_Crepeau_Robert_1986}), that is the variant where Alice prepares $n$ messages instead of two, and Bob gains full information about one of them (Alice does not know which one) and no information at all about the others.
	Quantum Oblivious Transfer (QOT) is the quantum version of this protocol, first introduced by Crépeau in \cite{quantum_oblivious_transfer_crepeau}. Unfortunately, with a proof similar to the one that proved the impossibility of unconditionally secure QBC, Lo proved that an unconditionally secure QOT is also impossible (\cite{lo_1997}).
	
	\subsection{One-Sided Two Party Computation (1S2PC)}
	\label{subsec:1s2pc} 
	One-Sided Two Party Computation (1S2PC) is an important cryptographic primitive that deals with the protection of private information during public decision. This is a protocol in which one party, Alice, wants to help the other party, Bob, compute the value of some function $f(j,k)$, where $j$, $k$ are private inputs given by Alice and Bob, respectively, that they do not want to reveal to the other party. More formally, a 1S2PC protocol is a protocol where Alice has a private input $j\in \{1,\dots,n\}$, Bob has a private input $k\in\{1,\dots, m\}$ and where Alice wants to help Bob computing the function $f(j,k)$. Then, the protocol is secure if:
	\begin{enumerate}
		\item Bob learns $f(j,k)$ unambiguously (for fixed values of $j$ and $k$);
		\item Alice learns nothing about $k$ or $f(j,k)$;
		\item Bob learns nothing about $j$.
	\end{enumerate}
	This class of protocols got involved in the chain of impossibility proofs in the late 1990s. In \cite{lo_1997}, Lo proved the impossibility of unconditionally secure 1S2PC, using essentially the same attack that proved the impossibility of unconditionally secure QBC. In fact, the proof is based on the fact that the previous three conditions cannot hold together. If Alice does not know anything about $k$, then Bob can cheat by applying a unitary transformation to his quantum machine and rotating from $f(j,k_1)$ to $f(j,k_2)$, thereby managing to obtain information about $f(j,k)$ for multiple values of $k$. This compromises the first requirement and also gives Bob the possibility to gain some information about the value of $j$.

	\subsection{Equivalence between QBC, QOT and 1S2PC}
	\label{subsec: qbcand1s2pc}
	We would like to highlight a few points in order to make the equivalence between QBC, QOT and 1S2PC protocols clear. First, we note that in \cite{yao_1995}, Yao proves that a secure QBC scheme can be used to implement a QOT protocol. Kilian, in \cite{Kilian_1988}, proves that a classical Oblivious Transfer protocol can be used to implement 1S2PC. It follows from these relations that the security of QBC implies the security of QOT, which implies the security of 1S2PC. 
	
	In order to prove the reverse implications, we should first agree that QOT is an example of 1S2PC. As proved by Lo in \cite{lo_1997}, we can reformulate it by saying that Alice inputs the pair of messages $j=(m_0,m_1)$ and Bob inputs $k\in \{0,1\}$, that is the index of the chosen message. At the end, Bob gains full information about the message he chose, that is $m_k=f(j,k)$. 
	According to this analysis, then one can conclude that the security of 1S2PC implies the security of QOT. Finally, following the protocol presented in \cite{bennett_brassard_crepeau_skubiszewska} by Bennett, Brassard, Crépeau and Skubiszewska, one can also show that the security of QOT would imply the security of QBC, hence closing the loop.

	\section{Quantum Private Queries (QPQ)}
	\label{sec:QPQ}
	Quantum Private Queries (QPQ), introduced in 2008 in \cite{Giovannetti_Lloyd_Maccone_2008}, is a protocol that addresses the Symmetric-Private Information Retrieval (SPIR) problem \cite{Gertner_Ishai_Kushilevitz_Malkin_2000}.
	There is a user, Alice, who wants to query a database, Bob. Suppose that the database has $n$ cells and Alice is interested in the $j$-th cell, $j\leq n $. Alice does not want to reveal to Bob which cell she is interested in, so a possible trivial solution for her would be to ask Bob for the whole database to ensure \textit{user privacy}. On the other hand, Bob does not want to disclose more information than is necessary to answer Alice's query. This requirement is called \textit{data privacy} and seems to conflict with user privacy.
	While QPQ does not offer an unconditionally secure solution to the SPIR problem, it guarantees 
	perfect data privacy and, relying on the no-cloning theorem and on the impossibility 
	to fully characterize a composite system by using only local operations,
	it permits Alice to implement  a cheat-sensitive test. The test can be passed by Bob with 
	probability $P=1$ if and only if he does not acquire information on Alice's query. In other words, QPQ ensures that: 
	\begin{enumerate}
		\item Alice learns unambiguously the value $A_j$ of the $j$-th data element; 
		\item Bob is guaranteed that Alice can learn at most two entries of the database (data privacy);
		\item if Bob tries to read $j$  or $A_j$, there is a non-zero probability $1-P>0$ he gets caught by Alice's cheating-test (if he chooses not to read $j$ or $A_j$ instead, he is sure to pass the test).  
	\end{enumerate}

	In the basic version of QPQ, Alice is required to prepare two registers with her queries: one contains her plain query, let say $\ket{j}_{\cal Q}$, while the other one contains a superposition of the query with a fixed record associated with a known answer, say $\frac{\ket{j}_{\cal Q}+\ket{0}_{\cal Q}}{\sqrt{2}}$. 
	She then sends them to Bob in random order. She waits for the response to the first one before sending the other one. Bob uses the qRAM algorithm \cite{Giovannetti_Lloyd_Maccone_2008b} to send the response. Let $\ket{A_j}_{\cal R}$ be the unique answer for the $j$-th query, then if Alice's query is $\ket{j}_{\mathcal Q}$, Bob sends back the registers $\ket{j}_{\mathcal Q}\otimes \ket{A_j}_{\mathcal R}$, while if Alice's query is $\frac{\ket{j}_{\mathcal Q}+\ket{0}_{\mathcal Q}}{\sqrt{2}}$, he sends back the entangled state 
	$|\Phi_{j}(A_j)\rangle_{\mathcal{Q,R}}:= \frac{\ket{j}_{\mathcal Q}\otimes\ket{A_j}_{\mathcal R}+\ket{0}_{\mathcal Q}\otimes\ket{A_0}_{\mathcal R}}{\sqrt{2}}$. So, there are two possible scenarios:
	\begin{itemize}
		\item Scenario $\ell=a$: Alice sends the plain query first and then the superposition. In such a case, if Bob is honest, the final state at the end of the protocol is of the form
		\begin{equation}
			\ket{\psi_{j}}^{(\ell=a)}=\left( \ket{j}_{\mathcal Q_1}\otimes \ket{A_j}_{\mathcal R_1} \right) \otimes 
			|\Phi_{j}(A_j)\rangle_{{\mathcal Q_2},{\mathcal R}_2}
			\, ,\label{equ1} 
		\end{equation}
		where $\mathcal Q_1$ and $\mathcal R_1$  represent the first query sent by Alice and the associated answer by Bob, while $\mathcal Q_2$ and $\mathcal R_2$ represent the second query and associated answer.
		\item Scenario $\ell=b$: Alice sends the superposition first. In this case, Eq.~(\ref{equ1}) gets replaced by:
		\begin{equation}
			\ket{\psi_{j}}^{(\ell=b)}=
			|\Phi_{j}(A_j)\rangle_{{\mathcal Q_1},{\mathcal R}_1} \otimes \left( \ket{j}_{\mathcal Q_2}\otimes \ket{A_j}_{\mathcal R_2}\right)
			\, ,\label{equ2} 
		\end{equation}
		where as in the case of scenario $\ell=a$, the couple $\mathcal Q_1$, $\mathcal R_1$ refers to the first query, and 
		$\mathcal Q_2$, $\mathcal R_2$ to the second one. 
	\end{itemize}
	In both scenarios, Alice can easily recover the value of $A_j$ by performing a simple von Neumann measurement on ${\cal R}_1$ (for $\ell=a$) or on ${\cal R}_2$ (for $\ell=b$). She can then use this result to run a test and determine whether the remaining registers contain the entangled state $|\Phi_{j}(A_j)\rangle_{\mathcal Q,R}$. 
	The security of the scheme then follows  from the fact that any attempts by Bob to recover the value of $j$ from registers ${\cal Q}_1$ and ${\cal Q}_2$ will result in deteriorations of such component which have a non-zero success probability $P$ of being  detected by Alice's test.

	\subsection{Probabilistic Quantum Private Queries (pQPQ) and Strong Probabilistic Quantum Private Queries (SpQPQ)}
	\label{subsec:pQPQ}

	An essential ingredient in the security proof of QPQ presented in~\cite{giovannetti_lloyd_maccone_2010} is that 
	Bob's database is deterministic, i.e., there is only one correct answer $A_j$ to each query $j$.
	More generally, one can consider 
	the probabilistic version of the problem obtained by assuming that  
	for each query $j$,   Bob's database  contains different correct answers $\{A_j^k\}_{k=1,\dots,m}$ which can be used to legitimately reply to Alice.
	Under these conditions one may ask whether it would be possible to replicate the results obtained for the deterministic database case, i.e., to device a probabilistic QPQ (pQPQ) algorithm that fulfills the following requirements:
	\begin{enumerate}
		\item Alice learns unambiguously the value $A^{k}_j$ of the $j$-th database element for a value of $k$ selected by Bob; 
		\item Bob is guaranteed that Alice can learn at most two entries of the database, say $A_j^{k}$ and $A_{j'}^{k'}$ (data privacy);
		\item  if Bob tries to read $j$  or $A_j^k$, there is a non-zero probability $1-P>0$ he gets caught by Alice's cheating-test (if he chooses not to read $j$ or $A_j^k$ instead, he is sure to pass the test).  
	\end{enumerate}
	It turns out that at least for the specific pQPQ design considered in Ref.~\cite{giovannetti_lloyd_maccone_2010}, if Bob is not committed to a particular value $k$, then 
	there is a special set of operations that he can perform which, while still ensuring points (1) and (2) of the above list, leads to an explicit violation of point (3), enabling him to pass Alice's cheating-test with probability $P=1$, even after having partially recovered  the value of $j$ (see Appendix~\ref{sec:Appendix} for details). 
	The question of whether this is a specific limit of the implementations analysed so far, or whether it is instead a consequence of a fundamental no-go theorem, is still an open problem. 
	Here we point out that a stronger version of the pQPQ problem (SpQPQ) obtained by imposing that Alice cannot
	recover $k$ from the received messages, and by replacing (3) with the request that Bob cannot have access to $j$,  is certainly not compatible with the structure of Quantum Mechanics. 
	In particular, in the next section we show that it is impossible to construct a SpQPQ protocol which realizes all the following tasks in an unconditionally secure way: 
	\begin{enumerate}
		\item Alice learns unambiguously the value $A^{k}_j$ of the $j$-th database element for a value of $k$ selected by Bob; 
		\item Bob is guaranteed that Alice can learn at most two entries of the database, say $A_j^{k}$ and $A_{j'}^{k'}$ (data privacy);
		\item Bob learns nothing about $j$ or $A_j^{k}$;
		\item Alice learns nothing about $k$.
	\end{enumerate}
	
	\section{Relations between SpQPQ and other protocols}
	\label{sec:relations}
	To show that an unconditionally secure SpQPQ protocol is impossible, in this section we prove that it
	is formally equivalent to 1S2PC. 
	
	It is easy to prove that the security of the 1S2PC protocol implies the security of SpQPQ. Indeed, SpQPQ is an example of 1S2PC if we rephrase it as follows: Bob wants to help Alice to compute a function $f(j,k)=A_j^k$, where $j$ is a private input of Alice (corresponding to the index of the query) and $k$ is Bob's private input, corresponding to the index he chooses among the possible correct answers. Then, if 1S2PC were unconditionally secure, at the end we would have that:
	\begin{enumerate}
		\item Alice learns $f(j,k)=A_j^k$ unambiguously (for fixed values of $j$ and $k$);
		\item Alice learns nothing about $k$;
		\item Bob learns nothing about $j$ or $f(j,k)$.
	\end{enumerate}
	This means that Alice receives the answer to her query and the protocol preserves user privacy.
	
	The vice-versa is immediate: if SpQPQ could be realized in an unconditionally secure way, then 1S2PC would be unconditionally secure too. In fact, if Bob wants to help Alice to compute a certain function $f(j,k)$, where $j$ is Alice's private input and $k$ is Bob's private input, then they could run an SpQPQ protocol. Alice would query for the $j$-th cell and Bob would answer with the response $A_j^k$. If SpQPQ were unconditionally secure, at the end of the protocol Alice would have learnt $A_j^k$ without having any information about the value of $k$, while Bob would have no information about $j$ nor $A_j^k$. Then, SpQPQ and 1S2PC are equivalent.
	
	Since 1S2PC is equivalent to both QOT and QBC, as recalled in Section \ref{subsec: qbcand1s2pc}, we can conclude that an equivalence holds among QBC, QOT, 1S2PC and SpQPQ protocols.

	\section{Conclusions}
	\label{sec:conclusions}
	In conclusion, we have formally defined two probabilistic versions of Quantum Private Queries protocol. The first one is the probabilistic Quantum Private Queries protocol (pQPQ), which was first  introduced in \cite{giovannetti_lloyd_maccone_2010}. Then, we have defined another version of this protocol with stronger requirements, namely, the Strong probabilistic Quantum Private Queries (SpQPQ), and investigated its security. We have shown that this protocol cannot be unconditionally secure by analyzing its relations with QBC, QOT and 1S2PC. Since these four protocols are equivalent, as shown in Section \ref{sec:relations}, and since the impossibility of unconditionally secure QBC, QOT and 1S2PC is well known (Sections \ref{subsec:qbc}, \ref{subsec: qot} and \ref{subsec:1s2pc}), it is clear that an unconditionally secure SpQPQ is also impossible. Even if a counterexample to the security of pQPQ has been shown in \cite{giovannetti_lloyd_maccone_2010}, the problem of formally proving its impossibility is still open. 
	\\
	
	The Authors acknowledge financial support by MUR (Ministero dell’Universit\`a e della Ricerca) through the PNRR MUR project PE0000023-NQSTI and project PRO3 Quantum Pathfinder. 
	
	\bibliography{biblio}

\begin{thebibliography}{10}

\bibitem{wiesner-1983}
Wiesner S.
\newblock Conjugate coding.
\newblock ACM SIGACT News. 1983;15(1):78–88.

\bibitem{Ardehali_1998}
Ardehali M. A simple quantum oblivious transfer protocol; 1998.
\newblock Available from: \url{https://arxiv.org/abs/quant-ph/9512026}.

\bibitem{bennett_brassard_crepeau_skubiszewska}
Bennett CH, Brassard G, Crépeau C, Skubiszewska MH.
\newblock Practical quantum oblivious transfer.
\newblock Advances in Cryptology — CRYPTO ’91;p. 351–366.

\bibitem{Brassard_Crepeau}
Brassard G, Crépeau C.
\newblock Quantum bit commitment and coin tossing protocols.
\newblock Advances in Cryptology-CRYPT0’ 90;p. 49–61.

\bibitem{Ardehali_1996}
Ardehali M. A quantum bit commitment protocol based on EPR states; 1996.
\newblock Available from: \url{https://arxiv.org/abs/quant-ph/9505019}.

\bibitem{brassard_crepeau_jozsa_langlois}
Brassard G, Crepeau C, Jozsa R, Langlois D.
\newblock A quantum bit commitment scheme provably unbreakable by both parties.
\newblock Proceedings of 1993 IEEE 34th Annual Foundations of Computer
  Science;.

\bibitem{Lo_Chau_1997}
Lo HK, Chau HF.
\newblock Is quantum bit commitment really possible?
\newblock Physical Review Letters. 1997;78(17):3410–3413.

\bibitem{Mayers_1996}
Mayers D. The trouble with Quantum Bit Commitment; 1996.
\newblock Available from: \url{https://arxiv.org/abs/quant-ph/9603015}.

\bibitem{lo_chau_1998}
Lo HK, Chau HF.
\newblock Why quantum bit commitment and ideal quantum coin tossing are
  impossible.
\newblock Physica D: Nonlinear Phenomena. 1998;120(1-2):177–187.

\bibitem{mayers_1997}
Mayers D.
\newblock Unconditionally secure quantum bit commitment is impossible.
\newblock Physical Review Letters. 1997;78(17):3414–3417.

\bibitem{lo_1997}
Lo HK.
\newblock Insecurity of quantum secure computations.
\newblock Physical Review A. 1997;56(2):1154–1162.

\bibitem{Giovannetti_Lloyd_Maccone_2008}
Giovannetti V, Lloyd S, Maccone L.
\newblock Quantum private queries.
\newblock Physical Review Letters. 2008;100(23).

\bibitem{giovannetti_lloyd_maccone_2010}
Giovannetti V, Lloyd S, Maccone L.
\newblock Quantum private queries: Security Analysis.
\newblock IEEE Transactions on Information Theory. 2010;56(7):3465–3477.

\bibitem{Rabin_OT}
Rabin MO.
\newblock How to exchange secrets with oblivious transfer.
\newblock Technical Report TR-81, Aiken Computation Lab, Harvard University.
  1981;.

\bibitem{Crepeau_1988}
Crépeau C.
\newblock Equivalence between two flavours of oblivious transfers.
\newblock Advances in Cryptology — CRYPTO ’87. 1988;p. 350–354.

\bibitem{Brassard_Crepeau_Robert_1986}
Brassard G, Crepeau C, Robert JM.
\newblock Information theoretic reductions among disclosure problems.
\newblock 27th Annual Symposium on Foundations of Computer Science (sfcs 1986).
  1986;.

\bibitem{quantum_oblivious_transfer_crepeau}
Crépeau C.
\newblock Quantum Oblivious Transfer.
\newblock Journal of Modern Optics. 1994;41(12):2445--2454.
\newblock Available from: \url{https://doi.org/10.1080/09500349414552291}.

\bibitem{yao_1995}
Yao ACC.
\newblock Security of quantum protocols against coherent measurements.
\newblock Proceedings of the twenty-seventh annual ACM symposium on Theory of
  computing - STOC '95. 1995;.

\bibitem{Kilian_1988}
Kilian J.
\newblock Founding cryptography on Oblivious transfer.
\newblock Proceedings of the twentieth annual ACM symposium on Theory of
  computing - STOC ’88. 1988;.

\bibitem{Gertner_Ishai_Kushilevitz_Malkin_2000}
Gertner Y, Ishai Y, Kushilevitz E, Malkin T.
\newblock Protecting data privacy in private information retrieval schemes.
\newblock Journal of Computer and System Sciences. 2000;60(3):592–629.

\bibitem{Giovannetti_Lloyd_Maccone_2008b}
Giovannetti V, Lloyd S, Maccone L. Quantum Random Access Memory; 2008.
\newblock Available from: \url{https://arxiv.org/abs/0708.1879}.

\end{thebibliography}
	\newpage
	\appendix
	\section{Counterexample about pQPQ}
	\label{sec:Appendix}
	We briefly recall the counterexample about the security of pQPQ provided in \cite{giovannetti_lloyd_maccone_2010}. This provides evidence for a strategy that Bob can use to retain some information about $j$ and still pass the honesty test. Suppose we have a probabilistic database with $N=3$ and where the correct answers for each $j$ are $A_0$ for $j=0$,  $A_1^{(\pm)}$ for $j=1$ and  $A_2^{(\pm)}$ for $j=2$.
	Suppose that, after Alice's first query, Bob replies using a  unitary transformation $U^{(1)}_{\mathcal{Q}_1, \mathcal{R}_1, \mathcal{B}}$ that induces the mappings 
	\begin{align*}
		\ket{0}_{\mathcal{Q}_1}\ket{0}_{\mathcal{R}_1}\ket{0}_{\mathcal{B}} &\to \ket{0}_{\mathcal{Q}_1}\ket{A_0}_{\mathcal{R}_1}\ket{0}_{\mathcal{B}}\;,  \\
		\ket{j}_{\mathcal{Q}_1}\ket{0}_{\mathcal{R}_1}\ket{0}_{\mathcal{B}} &\to \ket{j}_{\mathcal{Q}_1}\frac{|{A_j^{(+)}}\rangle_{\mathcal{R}_1}\ket{+j}_{\mathcal{B}} + |{A_j^{(-)}}\rangle_{\mathcal{R}_1}\ket{-j}_{\mathcal{B}} }{\sqrt{2}}\;, 
	\end{align*}
	where  $\ket{0}_{\mathcal{B}}, \ket{1}_{\mathcal{B}}, \ket{2}_{\mathcal{B}}$ are orthonormal states of local memory of Bob and where 
	for $j=1,2$, we set $\ket{\pm j}_{\mathcal{B}}:=\frac{\ket{0}_{\mathcal{B}} \pm \ket{j}_{\mathcal{B}}}{\sqrt{2}}$.
	After Alice's second query, instead Bob answers using a second unitary $U^{(2)}_{\mathcal{Q}_2, \mathcal{R}_2, \mathcal{B}}$  defined through the identities 
	\begin{align*}
		\ket{0}_{\mathcal{Q}_2}\ket{0}_{\mathcal{R}_2}\ket{\gamma}_{\mathcal{B}} &\to \ket{0}_{\mathcal{Q}_2}\ket{A_0}_{\mathcal{R}_2}\ket{\gamma}_{\mathcal{B}}\;, \\
		\ket{j}_{\mathcal{Q}_2}\ket{0}_{\mathcal{R}_2}\ket{\pm j}_{\mathcal{B}} &\to \ket{j}_{\mathcal{Q}_2}|{A_j^{(\pm)}}\rangle_{\mathcal{R}_2}\ket{\pm j}_{\mathcal{B}}\;,
	\end{align*}
	for all $\ket{\gamma}_{\mathcal{B}}$ of $\mathcal{B}$. Then, if Alice chooses $j=0$, the final state of the protocol is $\ket{0}_{\mathcal{Q}_1}\ket{A_0}_{\mathcal{R}_1}\ket{0}_{\mathcal{Q}_2}\ket{A_0}_{\mathcal{R}_2}\ket{0}_{\mathcal{B}}$. Bob passes the honesty test and gets $\ket{0}_{\mathcal{B}}$ on his private machine. If Alice chooses $j=1,2$, the final state of the protocol depends on the scenario: if $\ell=a$, then the final state is 
	\begin{equation}\nonumber 
		\frac{1}{\sqrt{2}} \; \left(\ket{j}_{\mathcal{Q}_1} \otimes |{A_j^{(+)}}\rangle_{\mathcal{R}_1}\otimes 
		|\Phi_j(A_j^{(+)})\rangle_{\mathcal{Q}_2\mathcal{R}_2} \otimes \ket{+j}_{\mathcal{B}} + 
		\ket{j}_{\mathcal{Q}_1} \otimes 
		|{A_j^{(-)}}\rangle_{\mathcal{R}_1}\otimes 
		|\Phi_j(A_j^{(-)})\rangle_{\mathcal{Q}_2\mathcal{R}_2} \otimes 
		\ket{-j}_{\mathcal{B}}\right)\;, 
	\end{equation}
	
	while if $\ell=b$ the final state is 
	\begin{equation}\nonumber 
		\frac{1}{\sqrt{2}}\left( 	|\Phi_j(A_j^{(+)})\rangle_{\mathcal{Q}_1\mathcal{R}_1}
		\otimes 
		\ket{j}_{\mathcal{Q}_2} \otimes  |{A_j^{(+)}}\rangle_{\mathcal{R}_2}\otimes \ket{+j}_{\mathcal{B}} +
		|\Phi_j(A_j^{(-)})\rangle_{\mathcal{Q}_1\mathcal{R}_1}
		\otimes 
		\ket{j}_{\mathcal{Q}_2} \otimes
		|{A_j^{(-)}}\rangle_{\mathcal{R}_2}\otimes \ket{-j}_{\mathcal{B}}\right) \,.
	\end{equation}
	Then, for each choice of $j$ and $\ell$, Bob passes the honesty test with certainty and Alice receives the answer $A_j^{(+)}$ half of the times and the answer $A_j^{(-)}$ half of the times. Moreover, when Alice gets  $A_j^{(+)}$, the state Bob holds in $\mathcal{B}$ is $|+j\rangle_{\cal B}$, while
	when Alice gets $A_j^{(-)}$, Bob has $|-j\rangle_{\cal B}$. Therefore, when Alice is querying the 
	index $j$, Bob in average gets the density matrix $(|0\rangle_{\cal B}\langle 0| + |j\rangle_{\cal B}\langle j|)/2$.
	This retains part of the information about $j$, which he can recover via a von Neumann measurement without getting caught by Alice. 
\end{document}